\documentclass{article}

\usepackage{arxiv}

\usepackage[utf8]{inputenc} 
\usepackage[T1]{fontenc}    
\usepackage{hyperref}       
\usepackage{url}            
\usepackage{booktabs}       
\usepackage{amsfonts}       
\usepackage{nicefrac}       
\usepackage{microtype}      
\usepackage{graphicx}
\usepackage{natbib}
\usepackage{doi}
\usepackage{amsmath}

\title{Composite metamorphic relations for integration testing}

\author{ \href{https://orcid.org/0000-0000-0000-0000}{\includegraphics[scale=0.06]{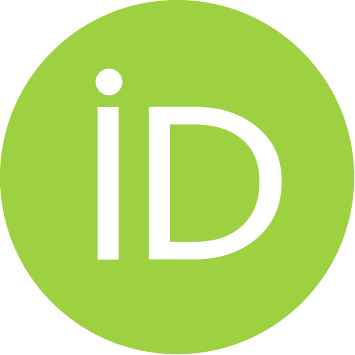}\hspace{1mm}Sofia F. Iakusheva} \\
	Moscow Institute of Physics and Technology\\
	9, Institutsky dr., Dolgoprudny, Moscow region, Russia \\
	\texttt{yakusheva.sf@phystech.edu} \\
	\And
	\href{https://orcid.org/0000-0000-0000-0000}{\includegraphics[scale=0.06]{orcid.pdf}\hspace{1mm}Anton S. Khritankov} \\
	HSE University\\
	Moscow, Russia \\
	\texttt{akhritankov@hse.ru} \\
}

\renewcommand{\shorttitle}{Composite metamorphic relations for integration testing}

\hypersetup{
pdftitle={Composite metamorphic relations for integration testing},
pdfsubject={q-bio.NC, q-bio.QM},
pdfauthor={Sofia Iakusheva, Anton Khritankov},
pdfkeywords={Software testing, metamorphic testing, integration testing, composite metamorphic relations, bioinformatics software testing.},
}

\begin{document}
\maketitle

\begin{abstract}
Metamorphic testing is a testing method for problems without test oracles. Integration testing allows for detecting errors in complex systems that may not be found during the testing of their components. In this paper, we propose a novel approach that applies metamorphic testing in integration testing. The main idea is to develop a composite metamorphic relation for the system represented as an acyclic graph. This relation is a logical function of metamorphic relations for the parts of the system (vertices of the graph). It takes into account the features of the parts. Also, it can simplify the search for failure by identifying the subsystem with error. In this paper's theoretical part, we describe an algorithm of relation design. Then, we apply our method to a bioinformatics system for comparative genetic analysis of tissues using production tools. This experiment proves our method can be applied to real-life pipelines and find errors in them.
\end{abstract}


\section{Introduction}
Nowadays, the number of components in scientific software is overgrowing. Pipes-and-filters (pipelines) are a widespread architectural style that integrates functional components into large systems. Pipelines are essential for many bioinformatics, machine learning, and computer vision applications. Unfortunately, they are hard to test because of their complexity. Although individual pipeline components are usually well tested in isolation, the integration testing of the whole pipeline may reveal unexpected errors.

Metamorphic testing is a popular method to test scientific software when test oracle is not readily available \cite{10.1145/3143561}. They also provide an easy automatization, as metamorphic relations allow for a more straightforward way to generate test sequences. Metamorphic testing can be combined with other methods and approaches. Application of metamorphic testing to integration testing has been studied in \cite{1579141}. 

Bioinformatics systems analyse big amounts of data with high precision. Even single-nucleotide mutation could be a reason for diseases such as cancer and Alzheimer's disease. So, the accuracy of genome analysis should be as high as possible. That is why it is important to test bioinformatics systems properly.

In this paper, we propose a novel method that benefits from the advantages of metamorphic relations in integration testing. First, we describe a mathematical model of the system and develop an algorithm to construct composite metamorphic relations from the relations for the system components. Then, we apply our method to a bioinformatics pipeline for mutation detection. Experiments proved our method to be viable and capable of detecting errors. 

This paper outlines as follows. In section 2, we study applications of metamorphic testing and approaches to combine metamorphic relations. Then in section 3, we describe the proposed method. Next, in section 4, we apply it to the bioinformatics pipeline. After that in section 5 we discuss other possible applications and generalizations of our approach. Finally, section 6 concludes the paper.

\section{Background and related work}

Software system for scientific analysis usually contains many components. Instruments for different steps of processing are often implemented by different authors. Pipes-and-filters is a popular architectural style for data processing because they easily combine those instruments into a complex system.

Examples of pipelines can be found among the bioinformatics systems, for example, in the mutation analysis. It is a popular area in bioinformatics, which can be applied for choosing cancer therapy. 

A mutation is an alteration in the nucleotide (base pair) sequence caused by different factors. Mutations can be somatic or germline. Somatic mutations are not inherited from parents and not passed to offspring. Germline mutation in the reproductive cells appears as a constitutional mutation in offspring. Also, mutations can be large-scale and small-scale. The term "indel" uses to refer to small insertions and deletions. Small-scale indels up to 50 base pairs are called microindels.

A typical bioinformatics pipeline consists of aligners and analyzing tools (variant callers, estimators of genes copy number, different tools for statistics calculation). A pipeline can contain dozens of such tools. Each component can be tested with metamorphic relations. However, integration testing is also required.

The main idea of metamorphic testing is to determine the presence of so-called metamorphic relations between all the inputs and outputs of the program. Those relations allow validating the program if the correct answers for each test are unknown. That automates the testing process and helps to generate large test sets. Metamorphic testing can be applied to autonomous driving systems, graph searching algorithms, search systems, query languages, and many others~\cite{10.1145/3143561}. Chan et al.~\cite{1579141} used metamorphic relations and checkpoints for integration testing. Tang et al.~\cite{7961646} and Giannoulatou et al.~\cite{Giannoulatou2014} applied metamorphic testing to popular bioinformatics tools. Troup et al.~\cite{10.1145/2896971.2896975} tested the cloud-based bioinformatics pipeline, but they consider the system in general.

It is possible to generate new metamorphic relations based on already known ones~\cite{https://doi.org/10.1002/stvr.437, 10.1145/3143561}. As~\cite{6319226} sows, the composition (complex function) of metamorphic relations can be more effective than individual relations by themselves. One way to create a new metamorphic relation is to use genetic algorithms~\cite{info10120392}. Following these ideas, we propose the method of combining metamorphic relations for integration testing.

\section{Proposed method} 

Let us denote $\overline{x}_i$ as a set of program inputs, $f(\overline{x}_i)$ as a set of program's outputs on the test $i$. The metamorphic for this program can be represented as
\begin{equation}\label{inv1}
R(\overline{x}_1, \overline{x}_2, \dots, \overline{x}_n, f(\overline{x}_1), f(\overline{x}_2), \dots, f(\overline{x}_n)) \longrightarrow \{0, 1\}, 
\end{equation}
where $n\ge2$ is the total number of test inputs.
We consider that a relation $R$ (\ref{inv1}) is defined on $dom(R)$.

If $\{\overline{x}_1, \overline{x}_2, \dots, \overline{x}_n\} \in dom(R)$, then $R = 1$ means that no errors were detected by the relation (true relation), and $R = 0$ means that there is an error (false relation). Outside $dom(R)$ the behavior of $R$~(\ref{inv1}) is undefined.

We consider that a software system can be represented by directed acyclic graph, in which vertices correspond to individual software components and edges denote data exchanges between them. Let $x_{i,j,k}$ denote input $1\le j\le m$ for vertex $i$ on test $k$. Than $f_{l,i,k}$ is output $l$ of vertex $i$ on test $k$ ($f_{l,i,k} = f_i(x_{i,1,k},\dots,x_{i,e_i,k})_l$). $x_{i,j,k}$ can either be output of a component or input of the whole system, $f_{i,j,k}$ be input of a component or output of the system. We also denote $\overline{x} \subseteq \{x_{i,j,k}\}$ and $f(\overline{x}) \subseteq \{f_{i,j,k}\}$.

We define the metamorphic relation for such system as a function
\begin{equation}\label{inv2}
\begin{split}
R(\overline{x}_1, \overline{x}_2, \dots, \overline{x}_n, f(\overline{x}_1), f(\overline{x}_2), \dots, f(\overline{x}_n))  = \\ R(x_{1,1,1}, \dots,x_{e_1,1,1}, \dots, x_{e_v,v,1}, \dots,x_{e_v,v,n}, \\ ~~~f_{1,1,1}, \dots,f_{r_1,1,1}, \dots, f_{r_v,v,1}, \dots,f_{r_v,v,n},
) \longrightarrow \{0, 1\},
\end{split}
\end{equation}
where $v$ is the number of vertices in the graph, $e_i$ is the number of input edges and $r_i$ is the number of output edges of $i$ vertex, $n$ is the number of tests.

\subsection{Composite metamorphic relations}
We define \emph{composite metamorphic relation} as a boolean function $C$ of metamorphic relations $R_{i,j}$~(\ref{inv1}) of components of complex system:
\begin{equation}\label{inv3}
    C(R_{1,1}, R_{1,2}, \dots, R_{1, m_1},\dots,R_{i, j}, \dots, R{_{v,m_v}}) \longrightarrow \{0,1\}
\end{equation}
 where $i=1,\dots,v$ is an index of a component (vertex in graph), $j=1,\dots,m_i$ is an index of individual metamorphic relation for the vertex $i$, $m_i$ is a number of chosen relations for this component. We calculate metamorphic relations for each vertex $R_{i,j}$ and then compute $C$~(\ref{inv3}). $C$ is a metamorphic relation for the whole system~(\ref{inv2}).

We define a subsystem as a subgraph that contains a subset $V \subseteq \{1,2,\dots,v\}$ of vertices and all existing edges between them.
Then the function
\begin{equation}
L(R_{1,1}, \dots, R_{1, m_1},\dots,R_{i, j}, \dots, R{_{v,m_v}}) =  L
 \begin{cases}
   1, &\text{$v_k \notin V$} \\ 
   R_{i,j}, &\text{$v_k \in V$} \\
 \end{cases} \longrightarrow \{0,1\}
\end{equation}
is a composite metamorphic relation for the subsystem.

If the relation is false for some vertex, then an error occurred either in it or in one of the vertices with a directed path to this vertex. That allows us to narrow down the search to a subsystem that contains the vertex. In this way, subsystems also could be tested using the same metamorphic relations as the whole system.

\subsection{Derivation of composite metamorphic relations}
For a system represented as an acyclic directed graph we present an algorithm of deriving composite metamorphic relations (\ref{inv3}) from metamorphic relations of the components (vertices).

Let us consider relations $A$, $B$, $R$ \eqref{inv1} for the system $f$ with set of inputs $x$. We define the function $def(z)$ that is false if $z$ is not known (or undefined) during the system execution and true if $z$ is known. Then, we define the following functions:

\begin{equation}
DEF(R, x) =
 \begin{cases}
   True, & x \notin dom(R)\\ 
   R(x), & x \in dom(R)
 \end{cases}
 \end{equation}
 
\begin{equation}
 INDEF(R, x)=
 \begin{cases}
   True, & \neg def(R(x))\\ 
   False, & def(R(x))
 \end{cases}
\end{equation}

\begin{equation}
A(x)\hat{\cup}B(x) =
 \begin{cases}
   A(x), &  x \notin dom(B)\\ 
   B(x), & x \notin dom(A) \\
   A(x) \cup B(x), & x \in dom(A)\cap dom(B) 
 \end{cases}
 \end{equation}
 
 \begin{equation}
 A(x)\hat{\cap}B(x) =
 \begin{cases}
   A(x), & x \notin dom(B))\\ 
   B(x), & x \notin dom(A) \\
   A(x) \cap B(x), & x \in dom(A)\cap dom(B) 
 \end{cases}
\end{equation}

 \begin{equation}
 A(x)\hat{\oplus}B(x) =
 \begin{cases}
   A(x), & x \notin dom(B)\\ 
   B(x), & x \notin dom(A)  \\
   A(x) \oplus B(x), & x \in dom(A)\cap dom(B) 
 \end{cases}
\end{equation}

Functions $A\hat{\cap}B$, $A\hat{\cap}B$ and $A\hat{\oplus}B$ have domains $dom(A) \cup dom(B)$.

Then, we propose an algorithm.

\textit{Input}. Software system represented as an acyclic directed graph. Sets of metamorphic relations $S_i = \{R_{i,j}\}$, $1 \le j \le m_i$ for each vertex $i$.

\textit{Output}. Composite metamorphic relations~\eqref{inv3} and their domains.

\begin{enumerate}

\item For each vertex, define set $S'_i$ of possible relations as a extension of $S_i$. Choose two relations $A$ and $B$ from $S_i$. Define relations $A \cup B$, $A \cap B$, $A \hat{\cup} B$, $A \hat{\cap} B$. Choose some of them that seem to be effective or easy-implemented and add them to $S'_i$. Do not add those with empty domains. Then, choose a new relation $C$ from $S_i$ and repeat the operation of combining with every element of $S'_i$. Repeat for every relation in $S_i$ and replace $S_i$ with $S'_i$.
\item Some relations in $S_i$ can be undefined during the execution. In this case, replace relation $R_{i,j}$ with relation $DEF(R_{i,j})$. If conditions when $R_{i,j}$ is not defined are known, divide $R_{i,j}$ into relations $R'_{i,j}$, $dom(R'_{i,j})\subset dom(R_{i,j})$, and $INDEF(R_{i,j})$ with the domain $dom(R_{i,j})/dom(R'_{i,j})$. Add both to $S_i$ and remove $R_{i,j}$.
\item For each vertex, choose single metamorphic relation $R_{i,j} \in S_i$. Let us name this set $U$. There can be many such sets.
\item For each set of relations $U_k$, replace $R_{i,j}$ with $DEF(R_{i,j})$ if vertex with relation $R_{i,j}$ uses the output of the vertex with $DEF$ or $INDEF$ relation.
\item Construct a complex metamorphic relation $C$ \eqref{inv3} for each set step by step. In the beginning, $C_0$ is a constant true.
\begin{itemize}
\item If vertex $i$ does not use the outputs of other vertices, or is the only one that uses the outputs of another single vertex $b$, then $C_{t}:=C_{t-1} \cap R_i$.
\item If many vertices use outputs of vertex $b$, define all possible functions of relations of these vertices with operations $\hat{\cup}$, $\hat{\cap}$, $\cup$ and $\cap$ (like in step 1). Choose one derived relation $A$ and define $C_t=C_{t-1}\cap A$. If there is another suitable relation $B$, duplicate the current set $U'=U$ and define $C'_t=C_{t-1} \cap B$ for it.
\end{itemize}

\item If there are such vertices $b_1$, $b_2$, ..., $b_n$ that has the same inputs and outputs,  and only one of them is computed during the test case (they are different execution branches),
\begin{enumerate}
    \item use operations $\hat{\cup}$, $\hat{\cap}$, $\hat{\oplus}$, $\cup$, $\cap$ and $\oplus$ to combine relations of these vertices into a single relation;
    \item or use these relations individually to acquire different composite metamorphic relations $C_1$, ..., $C_n$ and then combine these relations with the operations $\hat{\cup}$, $\hat{\cap}$, $\hat{\oplus}$, $\cup$, $\cap$ and $\oplus$ to get final composite metamorphic relation $C$.
\end{enumerate}
\item Choose composite relations with non-empty domains. If there are no such relations, choose other individual relations for the components  or other sets of relations for the system  (step 3) and repeat steps 4-7. 
\end{enumerate}


\subsection{Illustrative example}

\begin{figure}[h]
\centering
\includegraphics[width=0.6\linewidth]{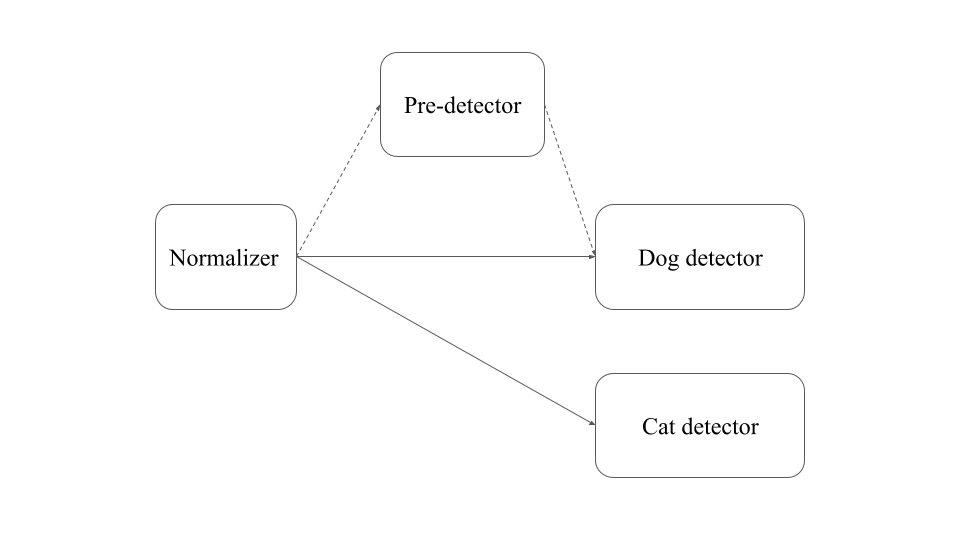}
\caption{System for illustrative example.}
\label{fig:model1}
\end{figure}

Let us demonstrate our algorithm on an example model of image detection system shown at Fig.~\ref{fig:model1}. It contains a cat detector, a dog detector with a pre-detector, and a normalizer, which is a piecewise linear function. Dog detector executes only when the output of pre-detector is True.

\subsubsection{Three-component example}

We consider this system with and without a pre-detector for better understanding.

Let us start with the system without a pre-detector. 

The normalizer is a piecewise linear function. It preserves the equivalence relation between pixels. If we add a cat to the picture without interaction with other animals, the cat detector should find all the animals it detected before and a new one. The same is for a dog detector if we add a dog to the picture.

Let us provide formal metamorphic relations for individual components on a pair of images $(a, b)$. 
The normalizer transforms those images into $(a' , b')$, than dog detector captures the dogs on those images as a set of bounding boxes $(B_d(a'), B_d(b'))$ and cat detector do the same for cats -- $(B_c(a'), B_c(b'))$.

\begin{description}
    \item[(N)] if there are equal pixels on $a$ and $b$ the same pixels should be equal on $a'$ and $b'$. 
    \item[(K)] If $b$ derives from $a$ by adding a cat without interaction with other animals $dom(K)$, the detector should find on $b$ all the cats it found on $a$, and a new one.
    \item[(D)] If $b$ derives from $a$ by adding a dog without interaction with other animals $dom(D)$, the detector should find on $b$ all the dogs it found on $a$, and a new one.
\end{description}

As we can see, $K$ is always false on $dom(D)$, and $D$ is always false for $dom(K)$. So we can extent relations $K$ and $D$ with $K^*$ and $D^*$.
 
\begin{description}
    \item[(K*)] If $b$ derives from $a$ by adding a cat without interaction with other animals $dom(K)$, the detector should find on $b$ all the cats it found on $a$ and a new one ($K$).
    If $b$ derives from $a$ by adding a dog without interaction with other animals $dom(D)$, the detector should find on $b$ all the cats it found on $a$ and not find the new ones.
    
    \item[(D*)] If $b$ derives from $a$ by adding a dog without interaction with other animals $dom(D)$, the detector should find on $b$ all the dogs it found on $a$, and a new one.
    If $b$ derives from $a$ by adding a cat without interaction with other animals $dom(K)$, the detector should find on $b$ all the dogs it found on $a$ and not find the new ones.
\end{description}


We construct composite relations for both sets $(N, K, D)$ and $(N, K^*, D^*)$ to analyze more possible cases. 

In the beginning, the composite relation $C_0$ is always true. Then $C_1=N$ because $N$ is an input vertex. $K$ and $D$ use its outputs. Their domains do not intersect, so we can use functions $K \hat{\cup} D$ and $K \hat{\cap} D$ (that should be equal in this particular case). Thus, we get relation $C=C_2=N \cap (K \hat{\cup} D)$.
For the other set, we can use functions $K^* \cup D^*$ and $K^* \cap D^*$ because their domains are the same. 
Thus, we get relations $C^*\in \{N \cap (K^* \cup D^*), N \cap (K^* \cap D^*)\}$.

Having established those relations, we can propose the test sequence generating rule. We can use pairs of images $(a, b)$, if $b$ is a copy of $a$ that contains one additional cat or dog.

\subsubsection{Four-component example}

Now let us move to a little bit more complex system shown at Fig.~\ref{fig:model1} that uses a pre-detector. Pre-detector has two different states - True and False. So we have execution branching, as shown in step 6. Let us consider that if True branch is taken the metamorphic relation $P$ holds, and if false branch is taken the relation is  $Q$. 

If we chose to combine relations for the pre-detector according to step 6.a, we can deduce relations 
$C=N \cap (P \cup Q) \cap (K \hat{\cup} D)$ and $C=N \cap (P \cup Q) \cap (K \hat{\cap} D)$.

If we chose to consider the system in different states  according to step 6.b, we can derive many relations like 
$C^1=N \cap ((P \cap(K \hat{\cup} D) \cup (Q \cap(K \hat{\cup} INDEF(D)))$, 
$C^2=N \cap ((P \cap(K \hat{\cap} D) \cup (Q \cap(K \hat{\cap} INDEF(D)))$,
$C^3=N \cap ((P \cap(K \hat{\cup} D) \oplus (Q \cap(K \hat{\cup} INDEF(D)))$.
The relation $C^4=N \cap ((P \cap(K \hat{\cap} D) \oplus (Q \cap(K \hat{\cap} INDEF(D)))$ seem to be the most strict.

Implementation of all described relations may be cumbersome because it is necessary to add animals without intersections.

\section{Application to a genetic analysis system}

Now let us derive composite metamorphic relations for a bioinformatics pipeline system. The experiment aims at answering the following research questions.

\textbf{RQ1.} Is it possible to derive and implement composite relations for the real-life system?

\textbf{RQ2.} Is it possible to detect failures using them?

\subsection{Composite metamorphic relation}

\begin{figure}[h]
\centering
\includegraphics[width=0.6\linewidth]{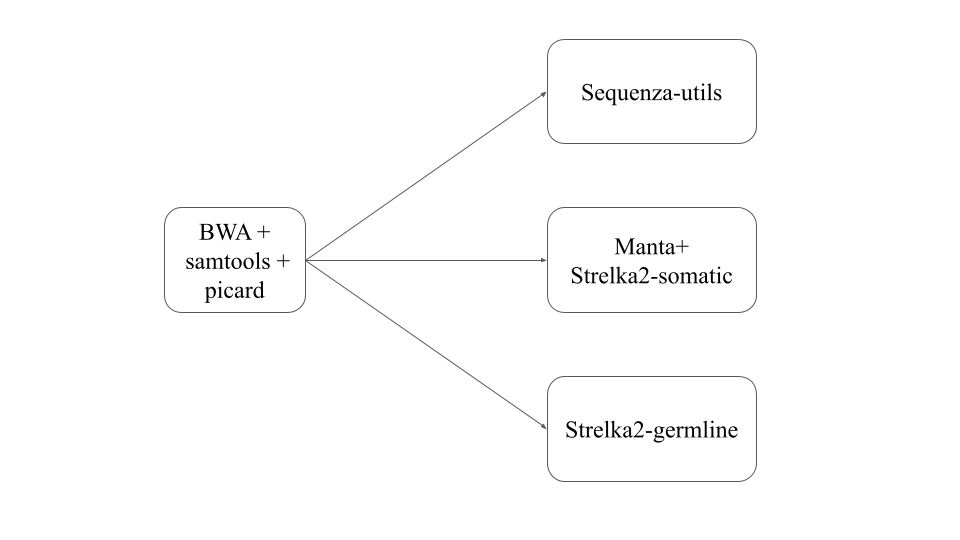}
\caption{System under test.}
\label{fig:model2}
\end{figure}

We chose a system for comparative genetic analysis of normal and tumor tissues. Such systems can contain several dozens of tools. We selected a subsystem that includes read aligner BWA \cite{li2013aligning}, variant caller Strelka2~\cite{strelka2}, supporting utility for statistics calculation Sequenza-utils~\cite{sequenza} and some auxiliary formatting tools.  We represent it as a directed graph shown at Fig.\ref{fig:model2}.

Input for this system is a standard reference genome and a pair of normal and tumor files in FASTQ formats. These files contain reads get with the sequencer and their parameters. BWA calculates a possible position of every read on the reference genome. Then, Strelka2-somatic searches for somatic microindels, Strelka2-germline searches for germline microindels, Sequenza-utils calculates statistics for copy number estimation. Strelka2-germline can process any FASTQ file, so it executes twice (on normal and tumor files) for better testing.

Let us follow the algorithm we describe in Section 3.2. The first step is to determine individual metamorphic relations for every vertex (pipeline component). 
 We use $(x_1,...,x_n)$ to denote inputs of each component and $f(x_1)$, ..., $f(x_n)$ to denote corresponding outputs for one test case.

\begin {description}
\item[BWA] ($\textit{BWA}$). If files $x_i$, $1\le i \le n$, contain mapped reads, all the outputs $f(x_i)$ should not be empty.
\item [Strelka2] ($S_i$, $GT_i$, $GN_i$). If $ x_i $ and $ x_{i+1} $ are two input files correspond $i$ and $i+1$ genomes, and $i+1$ genome is obtained from $i$ by adding independent mutations, then Strelka2 should find all the previous insertions and maybe new ones $ f (x_i) \subseteq f (x_{i+1}) $.
\item [Strelka2] ($S_d$, $GT_d$, $GN_d$). If $ x_i $ and $ x_{i+1} $ are two input files correspond $i$ and $i+1$ genomes, and $i+1$ genome is obtained from $i$ by adding independent mutations, then Strelka2 should find all the previous deletions and maybe new ones $ f (x_i) \subseteq f (x_{i+1}) $.
\item [Sequenza] ($SU$). If $ x_i $ and $ x_{i+1} $ are two input files correspond $i$ and $i+1$ genomes, and $i+1$ genome was obtained from $i$ by adding independent mutations, and Sequenza evaluates some of them by some metric, then those evaluations should reflect the changes correctly $ f (x_i) \le f (x_{i+1}) $ or $f (x_i) \ge f (x_{i+1})$ depending on the specific metric.
\end {description} 

 We define set of individual relations for functional parts of Strelka2. $GT$ denotes Strelka2-germline(tumor), $GN$ denotes Strelka2-germline(normal). We use relations $GT_i \cap GN_i$ and $GT_d \cap GN_d$. Also, $S$ denotes Strelka2-somatic(normal, tumor).

Mutations can be present in both normal and tumor samples or only in a tumor sample, thus they could be somatic or germline. They can not be germline and somatic at the same time, so we implemented the additional condition $Add$ (\ref{eq:germline}) for microindels.
\begin{equation}
\label{eq:germline}
    f_{germline}(tumor) = f_{germline}(normal) \oplus  f_{somatic}(normal, tumor)
\end{equation}

The next step is to define the domain of composite relation for sets $(\textit{BWA}, SU, S_i, GN_i, GT_i)$ and $(\textit{BWA}, SU, S_d, GN_d, GT_d)$. Firstly, $C_{i,1} = \textit{BWA}$ and $C_{d,1} = \textit{BWA}$. Three tools use BWA output. We choose relations $S_i \cap (GT_i \cap GN_i) \cap SU$ and $S_d \cap (GT_d \cap GN_d) \cap SU$. So, 

\begin{equation}
C_i := (\textit{BWA} \cap S_i \cap (GT_i \cap GN_i) \cap SU) \cap Add
,
C_d := (\textit{BWA} \cap S_d \cap (GT_d \cap GN_d) \cap SU) \cap Add
\end{equation}

Generated input files are based on biological genomes stored in files and represent normal and tumor tissues. We can use all possible inputs to test BWA if they are generated for genomes similar to the reference.
However, from these possible inputs, we can only use ones with sequentially added independent mutations to test Strelka2 and Sequenza. So, they are appropriate inputs for the system and the domain for the composite relations.

\subsection{Experiment configuration}

\begin{figure*}[h]
\centering
\includegraphics[width=0.85\textwidth]{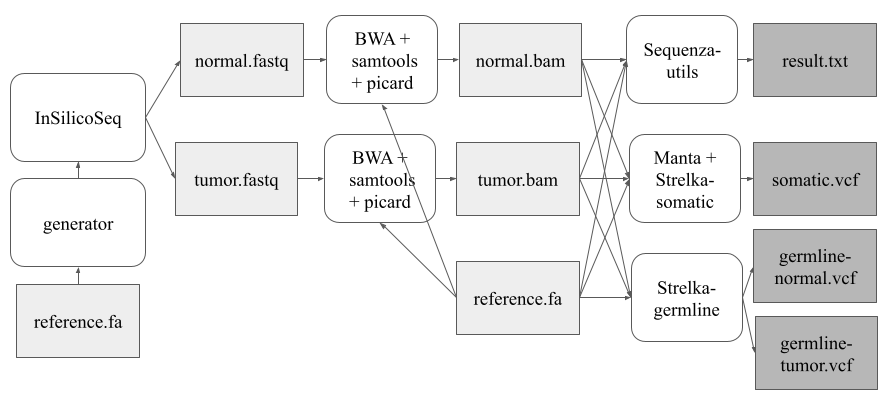}
\caption{The pipeline of the computational experiment. Test cases were composed by generator, then InSilicoSeq emulated sequencing process, and BWA, Strelka2, and Sequenza-utils analyzed produced data.} \label{fig:side}
\end{figure*}

We implement the proposed composite relation described in Section 4.2. Our generator shown at Fig.~\ref{fig:side} creates tests in two steps. First, it generates a configuration file that describes all mutations used in this test. It generates mutations randomly and uniformly with parameters described in the Table~\ref{params}. Then generator sequentially selects subsets of mutations and composes a configuration file for each test input. After that, it adds mutations into the reference genome and uses an instrument InSilicoSeq to generate data. Every test series contained 9 inputs.

Instrument InSilicoSeq generates data according to the probabilistic error model. This model emulates a real sequencing process which is very complicated. Generated reads can cover the genome non-uniformly and do not cover some random parts. Also, reads which cover ends of the genome are noisier. So, generated data often violate the relations. Thus, we test if the proposed relations can detect failures.

We generate two test groups. We add microindels-insertions in the first and microindels-deletions in the second, and we add large-scale insertions in both of them. We implement large-scale insertions as sequential duplications of the genome segment. All test cases are constructed according to the composite relation.

We use Carsonella ruddii genome \cite{carsonella} for the experiment. We divide it into disjoint segments and place microindels and large-scale insertions into them. Each segment contains mutations of one type to avoid interactions. We add germline mutations to both normal and tumor genomes and add somatic mutations to the tumor genome only.

We checked relations for the Strelka2 according to the values of the POS, REF, ALT columns in the output VCF file. We considered each file as a set of tuples (POS, REF, ALT), each tuple represents a mutation.  

We use number of detected failures as a metric 
\begin{equation}
\label{eq:failures}
    failures(i) = \frac{\sum\limits_{j=1}^{n}I(output(i,j)\not\subseteq output(i,j+1)}{n_i-1},
\end{equation}
where $output(i,j)$ is an $j$ test output for $i$ test case, $n_i$ is the number of test inputs in $i$ test case. This metric shows how many mutations can be not found during the subsequent test execution.

\begin{table}[h]
\centering
\caption{Execution parameters}
\begin{tabular}{llll}
\toprule
\textbf{Parameter} & \textbf{value} & \textbf{Parameter} & \textbf{value} \\
\midrule
Indel probability & 0.0005 &
Copynumber probability & 0.001 \\
Minimum indel size & 1 &
Maximum indel size & 50 \\
Minimum duplication size & 5000 &
Maximum duplication size & 10000 \\
\bottomrule
\end{tabular}
\label{params}
\end{table}

We check Sequenza-utils outputs visually. If the alteration appears on a certain position in a certain test, it should be present in the same position in all the remaining tests in this set. An example is shown at Fig.~\ref{fig:manmade}. For this purpose, we considered the output value 
\begin{equation}
    depth\_ratio (position)= \frac{N(tumor)}{N(normal)}
\end{equation}
where $N(tumor)$, $N(normal)$ are the number of reads that cover the considered position in the file with data for a normal sample and a tumor sample, respectively. If we place a duplication of some part of the genome, $depth\_ratio$ grows.

\subsection{Results} 
Tables~\ref{table2}  and~\ref{table3}  present the results of our experiment. 
Table~\ref{table2} represents the number of failed test cases and the number of violated individual metamorphic relations being parts of the composite ones ($BWA$ means BWA, $SU$ means Sequenza-utils,$GT$ means Strelka2 - germline (tumor), $GN$ means Strelka2 - germline (normal),  $S$ means Strelka2-somatic(normal, tumor), $Add$ means additional relation \eqref{eq:germline}. 

The experiment shows that all the composite relations are false. The input data contain errors because it is generated according to the sequencing model with noise. That means that the proposed composite relations are effective for error detection. According to Table~\ref{table2}, the composite relations seem to be more effective than the individual ones. 
\begin{table}[h]
\centering
\caption{The number of failed test cases for different experiment configurations, and the number of failed checks of individual relations in the composite one.}
\begin{tabular}{lccccccc}
\toprule
\textbf{}&\textbf{Failed tests} & \textbf{BWA} & \textbf{SU}& \textbf{GT} & \textbf{GN} & \textbf{S} & \textbf{Add}\\
\midrule
$C_i$ & 1.0 & 0.0 & 0.0 & 1.0 & 0.75 & 0.13 & 1.0\\
$C_d$ & 1.0 & 0.0 & 0.0 & 0.88 & 0.88 & 0.25 & 1.0\\
\bottomrule
\end{tabular}
\label{table2}
\end{table}

\begin{figure*}[h]
\centering
 \includegraphics[width=0.96\textwidth]{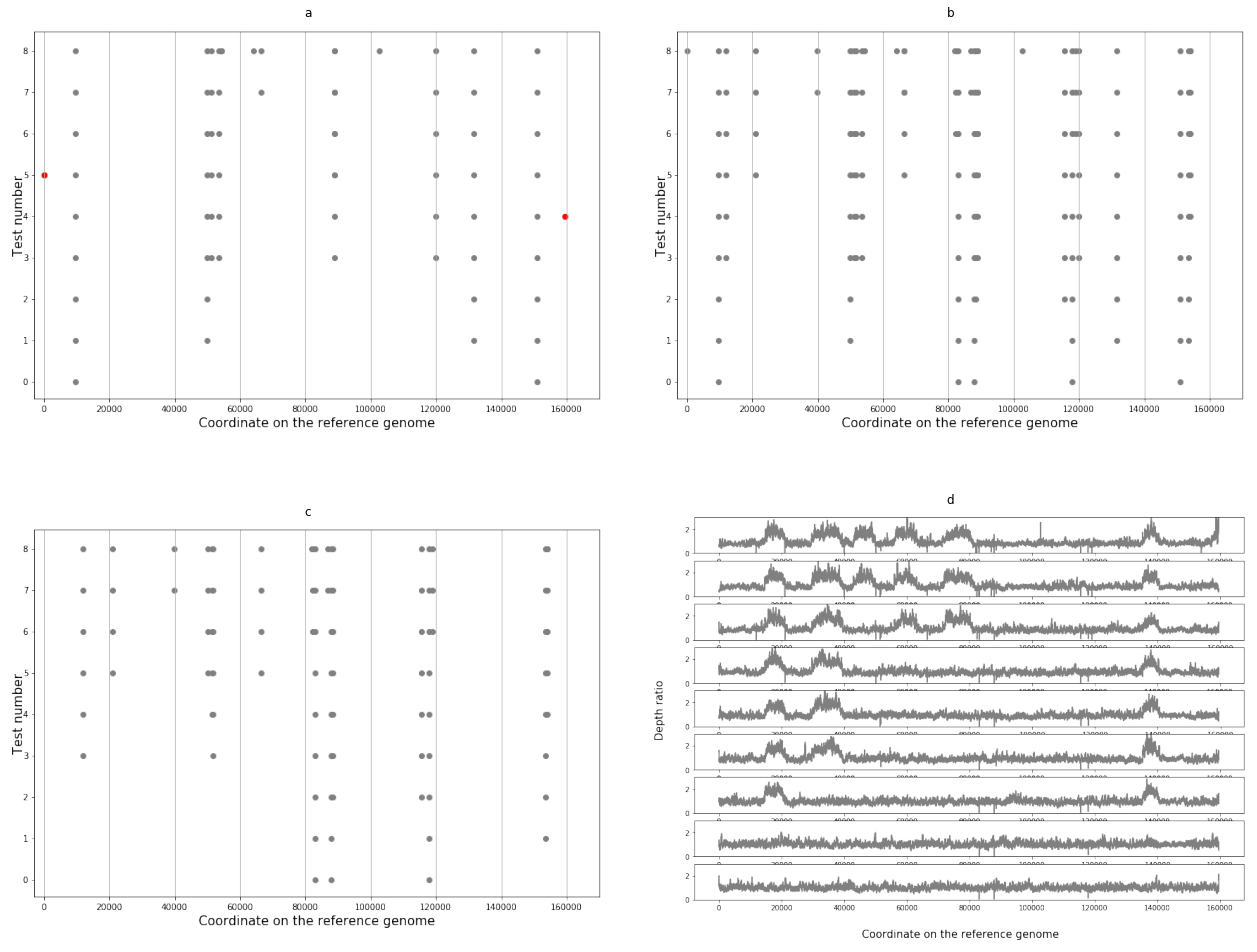}
\caption{Visualization of experiment results. }
\label{fig:manmade}
\end{figure*}

Figure~\ref{fig:manmade} contains a visualization of test outputs. Pictures a) and b) represent mutations found with Strelka2-germline: a) in normal genome compared to the reference, b) in tumor genome compared to the reference. Picture c) represents somatic mutations found with Strelka2-somatic. Picture d) represents copy number statistic. On a)-c) points represent indels coordinates on the reference genome. On d) spikes of the statistic mean found duplications. According to the proposed relations and corresponding generation of test sequences, if the mutation appears in a test, it should appear in all the remaining tests in the sequence. There are some red points in the picture a) which violate the relation. Appendix A contains more examples of false relations.

That confirms all the research questions.

Table \ref{table3} represents detected failures metric \eqref{eq:failures}. The results mean that proposed metamorphic relations for components are not always true. The degree of violating relations is different for the instruments. Also, the additional condition $Add$ seems to be more effective than other relations.

\begin{table}[h]
\centering
\caption{Metric \eqref{eq:failures} for different experiment configurations.}
\begin{tabular}{lcccc}
\toprule
\textbf{Configuration}& \textbf{GT} & \textbf{GN} & \textbf{S}&\textbf{Add}\\
\midrule
Microinsertions & 0.33 & 0.24 & 0.014 & 0.46 \\
Microdeletions & 0.35 & 0.32 & 0.028 & 0.47\\
\bottomrule
\end{tabular}
\label{table3}
\end{table}

\subsection{Threats to validity} 
The main internal threat to validity is wrong assumptions about bioinformatics tools and misinterpretation of their outputs. 
Another issue is that bioinformatics models are mostly probabilistic and usually have an acceptable error level. So, proposed metamorphic relations may not be suitable, for real-life testing. That is the direction of future work.

\section{Discussion and future work}

The proposed method combines metamorphic-unit and integration testing during the test execution. Integration testing often checks the most crucial system connections and compatibilities. Our approach can help to validate the inner system properties and structures. Also, it can help to minimize the number of executions and automatize integration testing.

For better testing automatization, it is necessary to develop novel methods for detecting potential metamorphic relations for components and composite relations for the whole system. There can be complex relations between many components that may not be connected directly, like an additional condition $Add$~\eqref{eq:germline}. 

Besides the pipelines, there are many other complex software systems in real-life applications, which should also be well-tested. So, an important issue is to develop integration testing methods for them. Moreover, it may be fruitful to combine metamorphic testing with other approaches.

The proposed method is focused on the relations between system components' inputs and outputs. However, it may also be helpful to check relations between the system states and inner parameters. For example, if the system contains a database, we can check its structure including schemas and data consistency during the execution.

A significant problem is not just to detect metamorphic relations but to select the most effective ones. So, the development of selection methods is a perspective and challenging area of research, as it seems to be complicated for huge systems with many different components.

Sometimes, composite relations may be cumbersome to construct and implement: systems can contain unstable or mutable components; the system structure can change during the execution, for example, if a system uses code-generation. The possible way is to calculate the metamorphic relations implicitly, for example, use asserts or checkpoints in the code. 

Of course, another interesting issue is the automatic construction of the composite metamorphic relations. For example, it can be convenient for testers to acquire test cases from the formal text description of the system. Model-checking, analysis of specifications, and some machine learning techniques may solve this problem. 

\section{Conclusion}

In this paper, we propose a novel approach for deriving composite metamorphic relations for a complex software system from relations of individual components. We implemented a composite metamorphic relation in practice to detect failures and tested a bioinformatics system. Our approach can be useful for testing scientific pipelines.

The composite relation allows checking many metamorphic relations with one test case. They can save time and resources. Proposed relations helped us detect and correct fails in experiment source code. Also, they were easily automated and useful to indicate misinterpretations and misunderstandings. 

Whereas in this paper, we focus on exact relation, the scientific systems may employ stochastic algorithms and allow for an appropriate error margin. Extension of the proposed approach to such relation could be a direction of future research.


\bibliographystyle{ACM-Reference-Format}

\appendix

\section{Failure examples}

\begin{figure*}[h]
\centering
 \includegraphics[width=0.9\textwidth]{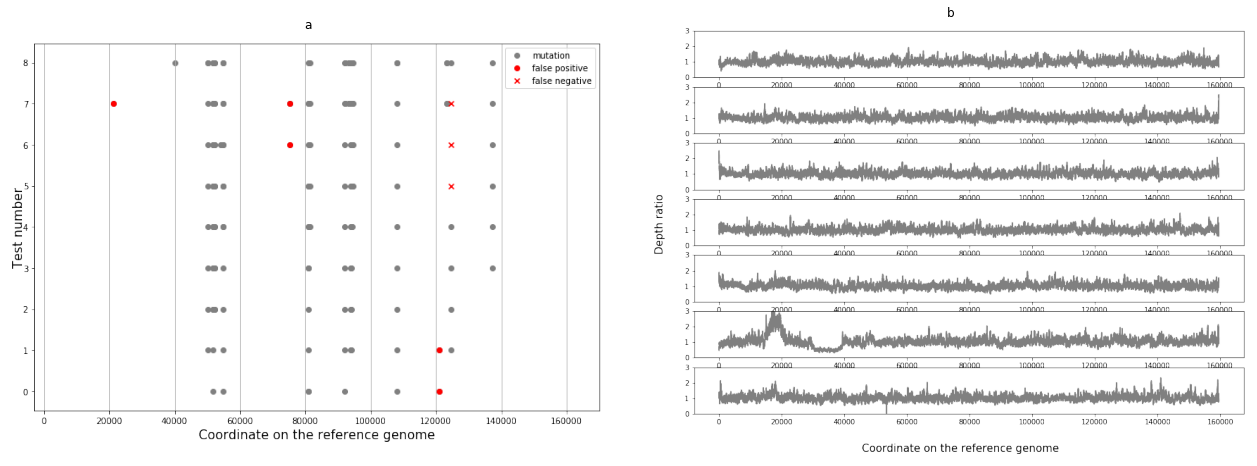}
\caption{Examples of false relations. }
\label{fig:failures}
\end{figure*}

We provide some examples for better understanding. Fig. \ref{fig:failures} shows the visualization of the experiments with heavy broken data. The proposed relations are false for them.

In the picture a), red points and crosses represent microindels that violate the individual relation $GN_i$. 

In picture b), the spike of the statistic appears in only one test output.

\end{document}